\documentclass[prd,twocolumn,showpacs,amsmath,amssymb,nofootinbib]{revtex4}

\newcommand\lsim{\lesssim}
\newcommand\gsim{\gtrsim}

\renewcommand\({\left(}
\renewcommand\){\right)}

\newcommand\eq[1]{Eq.~(\ref{#1})}
\newcommand\eqs[2]{Eqs.~(\ref{#1}) and (\ref{#2})}

\newcommand\eqreff[1]{(\ref{#1})}
\newcommand\eqsref[2]{(\ref{#1}) and (\ref{#2})}

\newcommand\ee{\end{equation}}
\newcommand\be{\begin{equation}}
\newcommand\eea{\end{eqnarray}}
\newcommand\bea{\begin{eqnarray}}

\newcommand\mpl{M_{\rm P}}

\newcommand\TeV{\,\mbox{TeV}}

\newcommand\mtwo{^{-2}}
\newcommand\mthree{^{-3}}

\begin{document}

\title{MSSM inflation}

\author{David H.~Lyth}

\affiliation{{\it Physics Department, Lancaster University, Lancaster
LA1 4YB, UK}}

\begin{abstract}
Variants of the $A$-term model of hep-ph/0605035 are considered. They are
equally successful, indicating that the model is quite robust once the relation
between $A$ and the soft mass is regarded as tunable.
 Alternatively a flat direction might support modular inflation.
\end{abstract}

\maketitle

\subsection*{1. The $A$-term model}

It has been pointed out recently \cite{tree} that a flat direction 
might support inflation, with a field value $\phi\ll\mpl$ and 
 the tree-level potential
\be
V =  \frac12 m^2\phi^2 - A 
\frac{\lambda_p \phi^p
}{
p\mpl^{p-3}
} + \lambda_p^2 
\frac{
\phi^{2(p-1)}
}{
\mpl^{2(p-3)}
} \label{pot}
.
\ee
This potential corresponds to soft supersymmetry breaking, 
keeping only a  term $W = \lambda_p \phi^p/\mpl^{p-3}$
in the superpotential with $p>3$.

Specifically, it is
suggested in \cite{tree} 
that we are dealing with one of the flat directions of the 
MSSM,  which provides two candidates  \cite{drt,gkm,flat} with $p=6$. 
This is a new
type of inflation model though, which might work with any flat direction
(gauge or singlet) 
that generates an $A$ term, and the  value of $p$
will not be specified in the following.

Slow-roll inflation  requires \cite{treview} 
the flatness conditions   $\epsilon\ll 1$,
$|\eta|\ll 1$ and $|\xi^2|\ll 1$ where\footnote
{Higher-order generalizations of $\xi^2$ should in general 
also be small but this is automatic in the cases we consider.
Near a maximum one can allow $|\eta|\sim 1$ if the inflaton
perturbation is not required to generate the curvature
perturbation.}
\bea 
\epsilon &=& \frac12 \mpl\mtwo \( \frac{V'}{3H_*^2} \)^2 \\
\eta &=& \frac{V''}{3H_*^2} \label{flat2}\\
\xi^2 &=& \frac{V'V'''}{9H_*^4} \label{flat3}
,
\eea
and $H_*$ is the inflationary Hubble parameter.

In the  small-field regime each of the three terms in \eq{pot}
(in fact any monomial) violates all of the flatness conditions
\cite{treview}. Two
terms might combine to generate a maximum but $\eta$ will still be too
big.  All three terms
must conspire to satisfy the conditions, and this is achieved if
\be
8(p-1)m^2 = A^2
 \label{aofm}
.
\ee
Indeed, this 
 gives $V'=V''=0$ at $\phi=\phi_0$, 
\be
\frac{\phi_0}{\mpl} =  \( \frac{m}{\lambda_p \sqrt{2p-2} \mpl}
\)^\frac1{p-2} \sim \( \frac m {\lambda_p \mpl} \)^\frac1{p-2}
\label{phi}
,
\ee
with 
\bea
V(\phi_0) &\sim & m^2 \phi_0^2 \label{v} \\
 V'''(\phi_0) &\sim& \frac{m^2}{\phi_0} \label{vppp} 
.
\eea

 In the regime
$|\phi-\phi_0| \ll |\phi_0|$ a good approximation is
\be
V = V(\phi_0) + \frac16  V'''(\phi_0) (\phi-\phi_0)^3
\label{vapp}
.
\ee
Slow-roll inflation ends when $|\eta|\sim 1$ at\footnote
{This is stronger than the condition $\eta\sim 1$ invoked
in \cite{tree}.}
\be
\frac{\phi_0-\phi}{\phi_0} \sim
 \( \frac{\phi_0}{\mpl} \)^3
,
\ee
which  is within the regime of validity of \eqreff{vapp}.

If $m^2$ is reduced slightly from the value \eqreff{aofm}, 
by an amount $\delta(m^2)$, the potential  becomes
\be
V = V(\phi_0) + \frac16  V'''(\phi_0) (\phi-\phi_0)^3
- \frac12\delta(m^2)  \phi^2
\label{vapp2}
.
\ee
There is now a maximum slightly below $\phi_0$, at which
$|\eta|
\sim \delta m^2/H_*^2$, where
$3\mpl^2H_*^2=V(\phi_0)$ is the
 Hubble parameter. Inflation can occur provided that
\be
\frac{\delta m}{m} \ll \frac{H_*}{m}
\label{deltam}
.
\ee
From \eqs{phi}{v}
\be
\frac {H_*}m  \sim \frac{\phi_0}{\mpl} \sim 
\( \frac  m {\lambda_p \mpl} \)^\frac1{p-2} \gsim  10^{-8}
\label{moverh}
.
\ee
The inequality comes from setting $m>\TeV$ and $p\geq 4$,
assuming $\lambda_p^\frac1{p-2} \sim 1$.

The condition \eqreff{aofm} is consistent with the expectation that
$m$ and $A$ have the same order of magnitude, but the accurate 
satisfaction of that condition implied by \eq{deltam} represents significant
fine-tuning. However, 
slow-roll inflation will presumably be preceded by an era of eternal inflation
with $\phi$ very close to the point where $V'=0$. If a landscape is available
with essentially continuous values for $m$ and $A$, this could justify the 
fine-tuning implied by \eq{deltam}. Alternatively one can just accept the
fine-tuning. 

Using the  potential \eqreff{vapp2}, the
 prediction for the spectrum of the curvature perturbation generated
by the inflaton perturbation depends on the parameters $m$ and $\delta m$
(with $A$ fixed at the value given by \eq{aofm}). Only the case $\delta m
=0$ has been  considered so far \cite{tree}, and the matter will not be pursued here.

\subsection*{2. Robustness of the  $A$-term model}

The flat inflationary potential is usually very sensitive to `corrections',
which can alter a model or prevent inflation altogether. Such
corrections can come from high-dimension terms in the tree-level potential
\cite{treview,km} and from loop corrections \cite{treview,mytev}.
As we now see, $A$-term inflation is essentially immune from these corrections.

Consider first  the likely value of  $\lambda_p$.
If the ultra-violet cutoff for the MSSM is $M\ll \mpl$,
 it is appropriate to replace $\mpl$ in \eq{pot} by $M$.
Then \cite{km} one may expect $\lambda_p\sim 1/p!$.
Retaining $\mpl$ in \eq{pot} these considerations give the estimate
\cite{km,ouraxion}
\be
\lambda_p \sim \( \frac\mpl{M} \)^{p-3} \frac1{p!}
\label{lamest}
.
\ee

If there is a GUT one may identify $M$ with the unification scale
$M\simeq 10\mtwo  \mpl$. Then \eq{lamest}  gives  $\lambda_6
\sim 10^3$ and $\lambda_9\sim 10^6$. If instead one takes $M\simeq \mpl$
the estimate becomes $\lambda_6\sim 10\mthree$ and $\lambda_9
\sim 10^{-6}$. 

A similar estimate will apply to higher-order  terms in $V$, of the 
form\footnote
{The most general possibility is $\Delta V\propto \phi^m$.
 This does not essentially change the following discussion.}
\be
\Delta V \sim \lambda_q^2  \frac{\phi^{2(q-1)}}{\mpl^{2(q-3)} }
,
\ee
with $q>p$.
Whatever the value of $M$, one presumably needs $\phi_0\lsim M$ 
for  such terms  to be under control. 

Given the condition $\phi_0\ll M$ it is reasonable to expect 
each higher-order term to give
 $\Delta V\ll V$. But one or more of them can still
have a significant effect on the very flat inflationary potential. 
Indeed,  the effect of $\Delta V$ on 
the flatness parameter $\eta$ (and $\xi^2$) is  
\be
\Delta \eta \sim  \( \frac{\lambda_q^2}{\lambda_p^2} \)
\( \frac\phi\mpl \)^{2(q-p-1)}
.
\ee
(To obtain this estimate we used $\Delta V''\sim \lambda_q^2 \phi^{2q-4}$
and $V\sim \lambda_p^2 \phi^{2p-2}$.)

At least if   $q = p+1$ one may expect $\Delta \eta\gsim 1$.
With the relation between $m$ and $A$ 
fixed by \eq{aofm}  this would  prevent inflation, but there
is no need to insist on this relation.
 We can make a small change  in $m^2$
(at fixed $A$) 
to recover again the condition $V'=V''=0$ at some point $\phi_0$.
Unless $\Delta \eta\lsim 1$ the change 
 will take $m^2$ outside  the original range 
\eqreff{deltam}, 
but that does not matter as long as we are taking the view that
the relation between $m$ and $A$ is to be specified in order to achieve 
inflation.
Provided that  $\Delta V\ll V$, the fractional change in $m^2$ will anyway
be small.

Now consider the one-loop correction to $V$.
 At the one-loop level
$m^2(\ln\phi)$ becomes a linear function of $\ln\phi$
\cite{treview}, with 
\be
\frac{
d\ln (m^2)
}{
d\ln \phi} \equiv \gamma \sim 
\alpha \tilde m^2/m^2 
.
\ee
Here  $\tilde m$ is the mass of the particle giving the dominant contribution
and $\alpha$ is its
coupling strength (gauge or Yukawa) to $\phi$. To justify the neglect of higher
loops one needs $|\gamma|\ll 1$.

The derivatives of the potential are now
\bea
V'/\phi &=& 
 m^2(1+\frac12\gamma) - A \lambda_p \phi^{p-2} \nonumber \\
&+&  2\lambda_p^2 (p-1) \phi^{2(p-2)}  \\
V'' &=&  m^2 (1+\frac32\gamma) - A\lambda_p (p-1) \phi^{p-2} \nonumber \\
&+& 2\lambda_p^2 (p-1)(2p-3) \phi^{2(p-2)} 
.
\eea
With $\gamma=0$, the requirement $V'=V''=0$ leads to the relation 
\eqreff{aofm} between $A$ and $m$ (and to \eq{phi} for $\phi_0$).
If we insist on that relation, the inclusion of $\gamma$ will spoil
inflation unless $\gamma$ is very small. But, analogously with the previous
discussion for a higher-order term in the tree-level potential, we can instead
make a  change in $m^2$, to achieve $V'=V''=0$ in the presence of $\gamma$.
Once the change has been made  the loop correction has a negligible 
effect, as a consequence of  the perturbative condition $|\gamma|\ll 1$.

Finally, there may be a rather different possibility for $A$-term inflation,
which is to use a $D$-flat but not $F$-flat direction. The relevant Yukawa
coupling might be many orders of magnitude below unity, as has been 
discussed in another context  \cite{kmt}, and then an $A$-term model
might be possible with $m^2\phi^2$ replaced by $\lambda\phi^4$. 

\subsection*{3. Parameter choices}

We focus now on a flat direction of the MSSM, corresponding to $p=6$.
To minimize the effect of multi-loop corrections,  
 $\phi_0$ should be roughly of order the renormalization scale
$Q$.
Given this identification there are at least two possible 
attitudes regarding the soft supersymmetry breaking parameters   $m$, $A$
and $\tilde m$.

As this   scale  is high ($\phi_0\gsim 10^{-4}\mpl$)
one can suppose  that the parameters 
are completely different from
 the corresponding quantities evaluated at the TeV scale,
which are  to be compared directly with observation. 
String   theory nowadays
offers so many possibilities that such a view is quite tenable,
but it  loses the direct connection with observation. 
Therefore it is more attractive to  suppose that the parameters are
those measured at the TeV scale.
This  view (assuming that supersymmetry is already broken at the 
high scale)  is perhaps supported by the 
 compatibility of the observed gauge couplings with
naive unification 

With ordinary supersymmetry, the 
parameters 
are  then $\alpha\sim 0.1$  and $m \sim \tilde m\sim \TeV$ 
(with  $\tilde m$ the gaugino mass). Keeping the gauge unification one
might  consider instead split supersymmetry \cite{split}, 
which makes the scalar masses very large while
keeping the gaugino masses of order $\TeV$. 
But the phenomenology then requires $A\ll m$  making the
$A$-term inflation unviable.

Throughout this discussion we have taken the view that the parameters
are to be specified directly at the scale $\phi_0$. Instead one might
suppose that they are defined at the GUT scale $M$.
In that case the parameters at the scale $\phi_0$ will have to be
calculated from the RGE's. To achieve the required relations 
\eqsref{aofm}{phi}
between $m^2$, $A$ and $\phi_0$ will require fine-tuning between the
values of $m^2$ and $A$ at the GUT scale, analogous to but not identical
with the relation \eq{aofm}.

\subsection*{4. Modular MSSM inflation}

Instead of invoking an $A$-term,  one can suppose that a
 flat direction  supports inflation with a 
 potential  of the form
\be
V(\phi) = V_0 f(\phi/\mpl)
\label{modpot}
,
\ee
where   $f$ and its low derivatives 
have magnitude of order 1 at a typical point  in the
regime $\phi/\lsim \mpl$. In the context of string theory, $\phi$ might
be a modulus with the vacuum a point of enhanced symmetry.

The  possibility that an MSSM  flat direction
could have the potential \eqreff{modpot}  has been mentioned before
(see for instance \cite{drt}) but 
 not  in connection with inflation.
Inflation with  the potential \eqreff{modpot} is indeed 
a quite attractive possibility
(see for instance \cite{al}).
Given the minimum at $\phi=0$, there will
typically be a maximum at $\phi\sim \mpl$, with $|\eta|\sim 1$.
Following the philosophy of modular inflation one either    hopes
 to get lucky so that actually $|\eta|\ll 1$, or else to  generate
the curvature perturbation through  a curvaton-type mechanism.

{\it Acknowledgments.}~ 
I thank Juan Garcia-Bellido, Anupam Mazumdar and Rouszbeh Allahverdi for
valuable comments on earlier versions of this paper, and  Fuminobu
Takahashi for pointing out the  relevance of \cite{kmt}.
 The  work was supported by PPARC grants  PPA/V/S/2003/00104,
PPA/G/O/2002/00098 and PPA/S/2002/00272 and by
 EU grant MRTN-CT-2004-503369.


\begin{thebibliography}{99}
\bibitem{tree}
 R.~Allahverdi, K.~Enqvist, J.~Garcia-Bellido and A.~Mazumdar,
  arXiv:hep-ph/0605035.
\bibitem{drt} M.~Dine, L.~Randall and S.~D.~Thomas,
  Nucl.\ Phys.\ B {\bf 458} (1996) 291.
\bibitem{gkm}
T.~Gherghetta, C.~F.~Kolda and S.~P.~Martin,
  Nucl.\ Phys.\ B {\bf 468} (1996) 37.
\bibitem{flat}
K.~Enqvist and A.~Mazumdar,
  Phys.\ Rept.\  {\bf 380} (2003) 99.
\bibitem{treview}
 D.~H.~Lyth and A.~Riotto,
  Phys.\ Rept.\  {\bf 314}, 1 (1999).
\bibitem{km}
C.~F.~Kolda and J.~March-Russell,
  Phys.\ Rev.\ D {\bf 60} (1999) 023504.
\bibitem{mytev}
D.~H.~Lyth,
  Phys.\ Lett.\ B {\bf 466} (1999) 85.
\bibitem{ouraxion}
 E.~J.~Chun, D.~Comelli and D.~H.~Lyth,
  Phys.\ Rev.\ D {\bf 62} (2000) 095013.
\bibitem{kmt}
S.~Kasuya, T.~Moroi and F.~Takahashi,
  Phys.\ Lett.\ B {\bf 593} (2004) 33.
\bibitem{split}
  N.~Arkani-Hamed and S.~Dimopoulos,
  JHEP {\bf 0506} (2005) 073.
\bibitem{al}
 L.~Alabidi and D.~H.~Lyth,
  arXiv:astro-ph/0510441.
\end{thebibliography}
\end{document}